\documentclass[sigplan,nonacm]{acmart}
\usepackage{fancyhdr}

\acmISBN{} 
\acmDOI{} 
\startPage{1}

\setcopyright{none}

\usepackage{booktabs}   
\usepackage{subcaption} 

\usepackage{listings}
\usepackage{soul}
\usepackage{color}
\makeatletter
\newcommand*{\whiten}[1]{\llap{\textcolor{white}{{\the\SOUL@token}}\hspace{#1pt}}}
\DeclareRobustCommand*\fancyunderline{%
    \def\SOUL@everyspace{\underline{\space}\kern\z@}%
    \def\SOUL@everytoken{%
     \setbox0=\hbox{\the\SOUL@token}%
     \ifdim\dp0>\z@
        \raisebox{\dp0}{\underline{\phantom{\the\SOUL@token}}}%
        \whiten{1}\whiten{0}%
        \whiten{-1}\whiten{-2}%
        \llap{\the\SOUL@token}%
     \else
        \underline{\the\SOUL@token}%
     \fi}%
\SOUL@}
\makeatother

\usepackage{mathpartir}

\usepackage{array}
\NewEnviron{syntax}{\footnotesize\begin{equation*}\begin{array}{r>{\hspace{-1em}}c<{\hspace{-1em}}ll}\BODY\end{array}\end{equation*}}
\newcommand{\termcolor}{\color[rgb]{0.5,0.1,0.1}}
\newcommand{\term}[1]{\text{\termcolor#1}}
\newcommand{\ntermcolor}{\color[rgb]{0.1,0.1,0.5}}
\newcommand{\nterm}[1]{\text{\ntermcolor\fancyunderline{#1}}}

\newcommand{\rulename}[1]{\textsc{#1}}

\NewEnviron{typerules}{\par\vspace{-\baselineskip}{\footnotesize\begin{align*}\BODY\end{align*}}}
\newcommand{\typerule}[3]{\left(\rulename{#1}\right)\text{ }&\inferrule{#2}{#3}&}

\NewEnviron{opsemrules}{\par\vspace{-\baselineskip}{\footnotesize\begin{align*}\BODY\end{align*}}}

\NewEnviron{typederiv}{\par\vspace{-\baselineskip}{\footnotesize\begin{equation*}\BODY\end{equation*}}}

\begin{document}

\title[]{Axon: A Language for Dynamic Shapes \\ in Deep Learning Graphs} 


\author{Alexander Collins}
\affiliation{
  \institution{NVIDIA}            
}
\email{acollins@nvidia.com}          

\author{Vinod Grover}
\affiliation{
  \institution{NVIDIA}           
}
\email{vgrover@nvidia.com}         

\begin{abstract}
Axon is a language that enables shape and rank inference for tensors in a Deep Learning graphs. It
aims to make shapes implicit and inferred, in a similar manner to how types are implicit and
inferred in many functional programming languages. Tensor dimensions are represented by expressions
consisting of symbolic variables, constants, and arithmetic operators. Tensor shapes can be
expressed as either a sequence of these dimension expressions, as a symbolic variable, or as an
appending of other shapes. This allows complex constraints on shapes to be expressed.

Axon is functional in style, with a type system similar in to Standard ML, extended to include shape
information. It provides a suite of built in operators over tensors, including pointwise arithmetic
operators, maps, reduction, loops and user defined functions.

We describe a shape inference algorithm based on constraint solving which infers information about
shapes, from both shape information provided by the programmer and the structure of the
program. This allows fully automatic inference of the shapes of tensors for complex Deep Learning
graphs.

This approach reduces programmer effort when specifying graphs, as tensor shapes are not explicit,
allows composition of Deep Learning graphs while maintaining input and output tensor shape
compatibility, and aids in automated error detection by identifying shape mismatches at runtime.

\end{abstract}

%


\maketitle

\thispagestyle{fancy}
\pagestyle{fancy}
\fancyhead{} 
\fancyfoot{} 
\fancyfoot[C]{\thepage}           

\section{Introduction}
\label{sec:introduction}

Deep Learning models can be viewed as constrained functional programs on tensor domains, which only
permit side effects or updates for certain types of models, and usually only during training.
Tensors have a type and a shape: they are rectangular domains of simple element types. Requiring the
programmer to deal with these shapes can add significant complexity to a language.

This paper describes shape and rank inference for tensors in a language we call Axon, which aims to
make shapes implicit and inferred, in a similar manner to how types are implicit and inferred in
many functional programming languages. Axon allows the individual dimensions of a tensor to be
expressions consisting of symbolic variables, constants, and arithmetic operators, allowing complex
constraints on shapes to be expressed. Furthermore, shapes can be expressed as either a sequence of
these dimension expressions, as a symbolic variable, or as an appending the dimensions of several
other shapes. This allows rank inference to be expressed using this shape appending.

An inference algorithm is also presented which infers information about shapes, from both shape
information provided by the programmer and the structure of the program, via a set of rules for
built-in operators. Our system allows the shapes involved in complex Deep Learning problems to be
automatically inferred without need for the programmer to express the shape constraints or give
concrete shapes which are potentially unknown until the shape of the inputs to the program are
known. Shape information can be inferred from known shapes, or partial shapes, of program inputs and
from the structure of the program itself. Automated error detection also aids programmers by
identifying shape mismatches between the allowed shapes to an operation and inferred shapes.

The rest of the paper is structured as follows. Section~\ref{sec:language} outlines the Axon
language. Section~\ref{sec:syntax} presents the syntax of shape expressions,
section~\ref{sec:constraint-generation} describes how standard Hindley-Milner type
inference~\cite{Hindley69,Milner78} is used to generate a set of shape constraints for the program
and section~\ref{sec:shape-solver} describes the algorithm used to solve sets of these shape
constraints. Section~\ref{sec:examples} provides examples of shape inference in action and
section~\ref{sec:example-graphs} provides larger examples for commonly seen Deep Learning
graphs. Section~\ref{sec:related-work} discusses related work and section~\ref{sec:conclusions}
presents our conclusions.

This paper makes the following contributions:
\begin{itemize}
\item
  A taxonomy of the kinds of shapes encountered in Deep Learning graphs, which we use to guide the
  design of the language.
\item
  A functional programming language, which we call Axon. It allows a programmer to specify symbolic
  shapes for input and output tensors for a graph, which permit arithmetic expressions for the
  dimensions, and allows for rank inference by expressing shapes as the composition of sub-shapes.
\item
  A constraint based solver for inferring shape information throughout a program, based on a set of
  shape inference rules. Initial constraints are generated based on the structure of a graph, and
  the rules are used to reduce these constraints to a fixed point, where either all shapes are known
  rank with constant dimensions, or are partially unknown allowing for runtime variable shape.
\end{itemize}

\section{The Axon Language}
\label{sec:language}

The goal of Axon is to provide a language for easily writing Deep Learning graphs, in a target
agnostic manner. It provides a high-level functional language for representing Deep Learning graphs,
through the use of a suite of built-in deep learning operators. The Axon compiler includes a type
inference algorithm, so that explicit types do not need to be specified by the
programmer. Furthermore, an inference algorithm for shapes also removes the need for explicit tensor
shapes to be specified by the programmer.

A simple graph that compute the pointwise sum and product of two inputs can be expressed in Axon as
follows:
\begin{lstlisting}
f = fn(x, y) {
  a = x + y;
  b = x * y;
  a, b
}
\end{lstlisting}

This program describes a graph called $f$ which takes two inputs $x$ and $y$ and produces two
outputs $a$ and $b$. In Axon, all variables are tensors. These have an element type (with the usual
primitive types including fixed size integers and floating point data types, and a shape.

Axon allows program decomposition into sub-graphs as in the following example:
\begin{lstlisting}
g = fn(a, b) {
  a + b
};
f = fn(a, b) {
  g(a, b) + a;
}
\end{lstlisting}
This allows a program to be decomposed into smaller modular parts and reused. The type and shape
inference algorithm allow for polymorphic types and shapes. For example, in the above graph, if $g$
were called from multiple places it could be called with different types and shapes.

Axon includes a suite of operators for performing Deep Learning computations. These include matrix
multiplication, convolution, point-wise arithmetic operators, loops and more. The language includes
a type specification for each built-in, which is used to infer types and shapes throughout the
graph. Examples of these are given in Section~\ref{sec:example-shape-constraints}.

\section{Tensor Shapes}
\label{sec:syntax}

In this section we describe the shapes of tensors in the Axon
language. Section~\ref{sec:taxonomy-of-shapes} outlines a general taxonomy of shapes we use to guide
our design and discussion. These are a general overview of the sorts of shapes that appear in Deep
Learning graphs. In Section~\ref{sec:syntax-of-shapes} we give the concrete syntax of shapes in the
Axon. Examples of shape constraints are given in Section~\ref{sec:example-shape-constraints}, which
show how shape constraints are generated from the syntactic structure of a graph. The algorithm for
then solving these constraints is given later, in Section~\ref{sec:shape-solver}

\subsection{Taxonomy of Shapes}
\label{sec:taxonomy-of-shapes}

Here we specify the general taxonomy of shapes we use to guide the design of our language. This
taxonomy categorizes the possible tensor shapes based on their properties and constraints.

\begin{itemize}
\item
  Concrete shapes

  These are shapes of statically known constant rank whose dimensions are all statically known
  constants.

  For example \lstinline|[]| denotes a scalar, \lstinline|[256]| denotes a one dimensional tensor of
  256 elements, and \lstinline|[1024, 256]| denotes a two dimension tensor.

\item
  Symbolic dimensions

  These extend concrete shapes by allowing individual dimensions to be symbolic expressions. The
  rank of the shape is still statically known.

  For example \lstinline|[x,y,16]| is a three dimensional tensor whose inner two dimensions are
  symbolic variables. At runtime, \lstinline|x| and \lstinline|y| will be known, but at compile time
  their value is unknown.

  \lstinline|[x*2,y,16]| is the shape that results from the concatenation of two tensors with the
  previous shape. Here the outermost dimension is a symbolic expression.
\item
  Symbolic

  These extend symbolic dimension shapes by allowing the for unknown rank.

  For example \lstinline|s @ [x,y]| denotes a tensor with at least two dimensions. \lstinline|s| is
  an arbitrary shape, and the \lstinline|@| operator is used to append \lstinline|s| to the shape
  \lstinline|[x,y]|.
\item
  Runtime dimensions

  Here the rank is static, but the individual dimensions can depend on the computation.
\item Runtime

  These extend runtime dimensions shapes by allowing the rank to also depend on computation.

\item
  Variable

  Dimension values differ across the tensor. This allows, for example, ragged batches to be
  represented.
\end{itemize}

\subsection{Syntax of Shapes}
\label{sec:syntax-of-shapes}

Figure~\ref{fig:shape-syntax} gives the syntax of shapes in Axon. The syntax of shapes is too
complex for a succinct grammar, therefore the grammar given in Figure~\ref{fig:shape-syntax}
generates some invalid shapes, which are further constrained by the shape construction rules given
in Figure~\ref{fig:shape-construction-rules}.

Shapes can be fixed, or variable, according to the dimension expressions they contain. The syntax of
these expressions includes positive integer constants, symbolic variables and arithmetic operations
and a ``variable dimension'' denoted \term{*}. The \term{*} dimension expression (with no arguments)
denotes a variable dimension, i.e. the number of elements in this dimension varies across the other
dimensions in the tensor. This allows, for example, batches of variable length sequences to be
represented. The shape of a batch of 10 variable length sequences of 3-dimensional vectors would be
denoted $\term{[10, *, 3]}$.

The shape append construct \term{@} allows for shapes to be expressed as many sub-shapes that are
appended to one another. This allows shape constraints such as $s \term{@} \term{[} n \term{]}$
which denotes a shape with a rank of at least one, where the outermost dimension has size
\term{n}. This is how Axon provides rank inference. We have found from our experience that this
append construct is able to represent the kinds of variable rank that we see in real Deep Learning
graphs.

\begin{figure}
  \begin{syntax}
    \nterm{Id}
       &=& [\nterm{Alpha} \enspace \term{\_}]
         [\nterm{Alpha} \enspace \nterm{Num} \enspace \term{\_}]^*
         & \text{Identifiers} \\
  \nterm{Alpha}
       &=& [\term{A}-\term{Z} \term{a}-\term{z}] \\
  \nterm{Num}
       &=& [\term{0}-\term{9}] \\
  \nterm{Shape}
       &=& \term{[]} & \text{Empty shape (scalar)} \\
    &\mid& \term{[} \nterm{Dimension} \big(\term{,} \enspace \nterm{Dimension}\big{)}^* \term{]}
         & \text{Shape with dimensions}\\
    &\mid& \nterm{Shape} \enspace \big( \term{@} \enspace \nterm{Shape} \big)^+
         & \text{Shape appending} \\
    &\mid& \nterm{Id} & \text{Shape variable} \\
  \nterm{Dimension}
       &=& \nterm{Num}^+ & \text{Dimension constant} \\
    &\mid& \term{*} & \text{Variable dimension} \\
    &\mid& \nterm{Id} & \text{Dimension variable} \\
    &\mid& \term{(} \term{+} \enspace \nterm{Dimension} \enspace \nterm{Dimension} \term{)}
         & \text{Addition} \\
    &\mid& \term{(} \term{-} \enspace \nterm{Dimension} \enspace \nterm{Dimension} \term{)}
         & \text{Subtraction} \\
    &\mid& \term{(} \term{*} \enspace \nterm{Dimension} \enspace \nterm{Dimension} \term{)}
         & \text{Multiplication} \\
    &\mid& \term{(} \term{/} \enspace \nterm{Dimension} \enspace \nterm{Dimension} \term{)}
         & \text{Division (floored)}
  \end{syntax}
  \caption{The syntax of shapes}
  \label{fig:shape-syntax}
\end{figure}

\begin{figure}
  \begin{typerules}
  \typerule{shape}
    {\forall i \in [1, n] \big( \vdash s_i : \nterm{Dimension} \big)}
    {\vdash \term{[} s_1 \term{,} \ldots \term{,} s_n \term{]} : \term{Shape}} \\
  \typerule{shape-append}
    {\forall i \in [1, n] \big( \vdash s_i : \nterm{Shape} \big)}
    {\vdash s_1 \enspace \term{@} \enspace \ldots \enspace \term{@} \enspace s_n : \term{Shape}} \\
  \typerule{shape-id}
    {s \in \nterm{Id}}
    {\vdash s : \term{Shape}} \\
  \typerule{dimension-constant}
    {d \in \nterm{Num}^+}
    {\vdash d : \term{Dimension}} \\
  \typerule{dimension-variable}
    { }
    {\vdash \term{*} : \term{Dimension}} \\
  \typerule{dimension-id}
    {d \in \nterm{Id}}
    {\vdash d : \term{Dimension}} \\
  \typerule{dimension-op}
    {\vdash d_1 : \term{Dimension} \and \vdash d_2 : \term{Dimension} \\ \and op \in \{+, -, *, /\}}
    {\vdash \term{(} op \enspace d_1 \enspace d_2 \term{)} : \term{Dimension}} \\
  \end{typerules}
  \caption{The rules for constructing shapes}
  \label{fig:shape-construction-rules}
\end{figure}

\subsection{Example Shape Constraints}
\label{sec:example-shape-constraints}

The following gives the types of commonly encountered deep learning operators. These type signatures
are used by the shape inference to infer information about the shapes of inputs and outputs and to
check that inputs and outputs to these operators are of a valid shape. When a graph is constructed,
rules such as these are used to create a set of shape constraints that can then be solved to infer
shape information about a graph. More details of how shape constraints are generated is given in
Section~\ref{sec:constraint-generation}, and the algorithm used to solve them is given in
Section~\ref{sec:shape-solver}.

\begin{itemize}
\item Concatenation \\
  $(\tau [d] @ s, \tau [d] @ s) \rightarrow \tau [(* \enspace d \enspace 2)] @ s$ \\
  This function takes two tensors which have shape $[d_1] @ s$ and $[d_2] @ s$, and concatenates
  them along their outermost dimension. This results in an output of shape $[(+ \enspace
    d_1 \enspace d_2)] @ s$. The outermost dimension of the output has a size which is equal to the
  sum of the outermost dimensions of the inputs, and the rest of the dimensions are arbitrary but
  match the inputs.
\item Matrix multiplication \\
  $(\tau s @ [d_1, d_2], \tau s @ [d_2, d_3]) \rightarrow \tau s @ [d_1, d_3]$ \\
  This function takes a tensor of shape $s @ [d_1, d_2]$ and a tensor of shape $s @ [d_1, d_2]$ and
  produces an output tensor of shape $s @ [d_1, d_3]$. $s$ is the shape of the batch (i.e. the
  matrix multiplication operates on the outer two dimensions of the inputs, and the outer dimensions
  must be equal, and are denoted $s$). The outer dimensions of the output shape are $d_1, d_3$,
  which expresses the fact that the result has the shape of the first dimension of the first input,
  and the second dimension of the second input. Finally, the inputs share dimension expression $d_2$
  which expresses the need for the outer and inner dimensions of the two matrices to be equal.
\item 2D Convolution in NCHW format \\
  $(\tau [n,c,h,w], \tau [k,c,r,s]) \rightarrow \tau [n, c, 1 + h - r, 1 + w - s]$ \\
  The shape of the output of a convolution is an arithmetic function of the shape of the inputs.
\end{itemize}

\pagebreak

\section{Shape Constraint Generation}
\label{sec:constraint-generation}

Our approach extends Hindley-Milner type inference~\cite{Damas1982TypeInference}, using standard
type unification with an additional rule which generates shape constraints. This rule is described
in Section~\ref{sec:constraint-generation-rules}. From these generated shape constraints, a shape
solver (based on unification) computes concrete shapes where possible from these shape
constraints. This part of the algorithm is covered in detail in Section~\ref{sec:shape-solver}.

Given a program, which is functional in style and may be annotated with shapes by the programmer,
Hindley-Milner type inference with standard unification is run over the program in order to infer
the types of all functions, values and let bindings. During this process, the additional rule in the
unification algorithm generates a set of shape constraints for the program. A shape solver is run
over these constraints in order to solve them. Any shape variable assignments that the shape solver
discovers are then replaced in the types of the program. This produces programs with concrete shape
information where enough information is present to deduce them.

\subsection{Constraint Generation Rules}
\label{sec:constraint-generation-rules}

In order to infer information about shapes within a program, a set of shape constraints first needs
to be generated for a program. This is done as part of the standard Hindley Milner style type
inference and unification algorithms.

Unification is augmented to generate not only a type substitution for a pair of types, but also a
set of type constraints. Whenever two tensor types involving shapes are unified, shape constraints
are generated and added to the set. After unification is complete these shape constraint sets are
solved using the shape solving algorithm presented in section~\ref{sec:shape-solver}. This shape
constraint generation and solving is done incrementally as part of the type inference algorithm.

The additional unification rules for tensor types are given in
Figure~\ref{fig:unification-tensor-rules}. $G$ denotes the set of types to be unified (initially $G =
\{ \tau_1 = \tau_2 \}$ where $\tau_1$ and $\tau_2$ are the two types being unified. $S$ denotes the
set of shape constraints, initially $S = \emptyset$. If any rules match $\langle G, S \rangle$ and
generate a $\text{FAIL}$ then unification fails. Unification finishes when none of the rules match
anything in $\langle G, S \rangle$. At this point, $G$ will contain a substitution from type
variables to types, and $S$ will contain the set of shape constraints.

The first rule in Figure~\ref{fig:unification-tensor-rules} just reorders the type so that type
variables (denoted $a$) appear on the left hand side. The second rule generates a new shape
constraint given two tensor types with shapes $s_1$ and $s_2$. The third rule causes unification to
fail if a free type variable appears on both the left hand side (denoted $a$) or anywhere in the
type on the left hand side (denoted $\tau$).

\begin{figure}
  \begin{align*}
    \langle G \cup \{ \tau s = a \}, S \rangle
    & \Rightarrow \langle G \cup \{ a = \tau s \}, S \rangle \\
    \langle G \cup \{ \tau_1 s_1 = \tau_2 s_2 \}, S \rangle
    & \Rightarrow \langle G \cup \{ \tau_1 = \tau_2 \}, S \cup \{ s_1 = s_2 \} \rangle \\
    \langle G \cup \{ a = \tau s \}, S \rangle
    & \Rightarrow \text{FAIL} \enspace \text{if} \enspace a \in \text{ftv}(\tau)
  \end{align*}
  \caption{Unification rules for tensor types to generate shape constraints}
  \label{fig:unification-tensor-rules}
\end{figure}

\section{Shape Solver}
\label{sec:shape-solver}

Type inference, as described in Section~\ref{sec:constraint-generation} produces a set of shape
constraints of the form:
\begin{equation*}
  C = \{ \dot{s}_1 = \dot{s}'_1, \ldots, \ldots \dot{s}'_n = \dot{s}'_n \}
\end{equation*}

The shape solver attempts to simplify this constraint set, and deduce concrete values for shapes
where possible. As shapes can depend on runtime values, the shape solver may not be able to
statically determine the value for all shape expressions. The solver uses an algorithm similar to
unification.

The $\text{solve-shapes}$ algorithm applies the following rules to the set of constraints until a
fixed point is reached. Rules are matched on the syntax of shapes, where $a$ denote shape variables,
$c$ denote integer constants, $s$ denotes shapes, $d$ denotes dimensions. $\dot{s}$ denotes both
shapes and dimensions. Two shapes are equivalent ($\equiv$) if their syntax is identical. If
\text{FAIL} is reached, the shape solver fails (as an invalid constraint has been found).

$C$ is the constraint set to be reduced. $C[\dot{s}/a]$ denotes the constraint set with all $a$
replaced with $\dot{s}$. $\text{fsv}(\dot{s})$ denotes the set of shape and dimension variables that
are free in $\dot{s}$.

\subsection*{Basic Rules}
These rules remove constraints that are tautologies, perform replacement of free shape variables
that have been determined and emit an error when different constants are equated to each other in a
constraint.
\begin{align*}
  C \cup \{ \dot{s}_1 = \dot{s}_2 \}
  & \Rightarrow C \enspace \text{if} \enspace \dot{s}_1 \equiv \dot{s}_2 \\
  C \cup \{ a = \dot{s} \}
  & \Rightarrow C[\dot{s}/a] \cup \{ a = \dot{s} \} \\
  & \qquad \text{if} \enspace a \in \text{fsv}(C) \enspace \text{and} \enspace
    a \notin \text{fsv}(\dot{s}) \\
  C \cup \{ c_1 = c_2 \}
  & \Rightarrow \text{FAIL} \enspace \text{if} \enspace c_1 \not\equiv c_2
\end{align*}

\subsection*{Reordering}
These rules reorder constraints with a constant on one side so that it is on the right hand
side. These rules remove the need for many symmetric rules later.
\begin{align*}
  C \cup \{ c = \dot{s} \}
  & \Rightarrow C \cup \{ \dot{s} = c \} \\
  C \cup \{ c = a \}
  & \Rightarrow C \cup \{ a = c \} \\
  C \cup \{ \term{*} = a  \}
  & \Rightarrow C \cup \{ a = \term{*} \} \\
  C \cup \{ \term{(op} \enspace d_1 \enspace d_2 \term{)} = a \}
  & \Rightarrow C \cup \{ a = \term{(op} \enspace d_1 \enspace d_2 \term{)} \} \\
  C \cup \{ \term{[} d_1 \term{,} \ldots \term{,} d_n \term{]} = a \}
  & \Rightarrow C \cup \{ a = \term{[} d_1 \term{,} \ldots \term{,} d_n \term{]} \} \\
  C \cup \{ s_1 \enspace \term{@} \enspace \ldots \enspace \term{@} \enspace s_n = a \}
  & \Rightarrow C \cup \{ a = s_1 \enspace \term{@} \enspace \ldots \enspace \term{@} \enspace s_n \} \\
\end{align*}

\subsection*{Unpacking}
These rules equate the shape expressions making up shapes of known rank.
\begin{align*}
  & C \cup \{ \term{[} d_1, \ldots \term{,} d_n \term{]} = \term{[} d'_1, \ldots \term{,} d'_n\term{]}
  \} \\
  & \Rightarrow C \cup \{ d_1 = d'_1, \ldots, d_n = d'_n \} \\
  & C \cup \{ \term{[} d_1, \ldots \term{,} d_n \term{]} = \term{[} d'_1, \ldots \term{,} d'_m
    \term{]} \} \\
  & \Rightarrow \text{FAIL} \enspace \text{if} \enspace n \neq m
\end{align*}

\subsection*{Empty shapes}
These rules assign the empty shape to shapes appended either side of a shape of known rank, where
possible.
\begin{align*}
  & C \cup \{ \term{[} d_1, \ldots \term{,} d_n \term{]} = s_1 \enspace \term{@} \ldots
  \term{@} \enspace s_j \enspace \term{@} \enspace \term{[} d'_1, \ldots \term{,} d'_n \term{]} \\
  & \enspace\enspace\enspace \term{@} \enspace s_{j+1} \enspace \term{@} \ldots \term{@} \enspace s_m \} \\
  & \Rightarrow C \cup \{ \term{[} d_1, \ldots \term{,} d_n \term{]} = \term{[} d'_1, \ldots \term{,}
  d'_n \term{]}, s_1 = \term{[]}, \ldots, s_m = \term{[]} \} \\
  & C \cup \{ s_1 \enspace \term{@} \ldots \term{@} \enspace s_j \enspace \term{@} \enspace
  \term{[} d'_1, \ldots \term{,} d'_n \term{]} \enspace \\
  & \enspace\enspace\enspace \term{@} \enspace s_{j+1} \enspace \term{@} \ldots \term{@} \enspace s_m = \term{[} d_1, \ldots \term{,} d_n \term{]} \} \\
  & \Rightarrow C \cup \{ \term{[} d_1, \ldots \term{,} d_n \term{]} = \term{[} d'_1, \ldots \term{,} d'_n \term{]},
  s_1 = \term{[]}, \ldots, s_m = \term{[]} \}
\end{align*}

\subsection*{Append Unpacking}
These rules equate dimension expressions that are at the start and end of appended shapes.
\begin{align*}
  & C \cup \{ s \enspace \term{@} \enspace \term{[} d_1 \term{,} \ldots d_n \term{]} =
    s' \enspace \term{@} \enspace \term{[} d'_1 \term{,} \ldots d'_m \term{]} \} \\
  & \Rightarrow C \cup \{ s \enspace \term{@} \enspace \term{[} d_1 \term{,} \ldots d_{n-1} \term{]} =
    s' \enspace \term{@} \enspace \term{[} d'_1 \term{,} \ldots d'_{m-1} \term{]}, d_n = d'_m \} \\
 & C \cup \{ \term{[} d_1 \term{,} \ldots d_n \term{]} \enspace \term{@} \enspace s =
    \term{[} d'_1 \term{,} \ldots d'_m \term{]} \enspace \term{@} \enspace s' \} \\
  & \Rightarrow C \cup \{ \term{[} d_2 \term{,} \ldots d_n \term{]} \enspace \term{@} \enspace s =
    \term{[} d'_2 \term{,} \ldots d'_m \term{]} \enspace \term{@} \enspace s', d_1 = d'_1  \}
\end{align*}

\subsection*{Append Dimensionality}
These rules check for shape appends whose minimum rank is greater than a concrete shape that they
are equal to.
\begin{align*}
  & C \cup \{ s_1 \enspace \term{@} \enspace \ldots \enspace \term{@} \enspace s_n =
    \term{[} d_1 \term{,} \ldots \term{,} d_m \term{]} \} \\
  & \Rightarrow \text{FAIL} \enspace \text{if number of dimensions in $s_1, \ldots, s_n > m$ } \\
  & C \cup \{ \term{[} d_1 \term{,} \ldots \term{,} d_m \term{]} =
    s_1 \enspace \term{@} \enspace \ldots \enspace \term{@} \enspace s_n \} \\
  & \Rightarrow \text{FAIL} \enspace \text{if number of dimensions in $s_1, \ldots, s_n > m$ }
\end{align*}



\subsection*{Simplification}
These rules use the $\text{simp}$ function to simplify shapes in the constraints.
\begin{align*}
  & C \cup \{ \dot{s}_1 = \dot{s}_2 \} \\
  & \Rightarrow C \cup \{ \dot{s}'_1 = \dot{s}'_2 \} \\
  & \qquad \text{where} \enspace
      \dot{s}'_1 = \text{simp}(\dot{s}_1), \enspace
      \dot{s}'_2 = \text{simp}(\dot{s}_2) \enspace \\
  & \qquad \text{if} \enspace
      \dot{s}_1 \neq \dot{s}_1' \enspace \text{or} \enspace \dot{s}_2 \neq \dot{s}'_2 \\
\end{align*}

The $\text{simp}$ function takes a shape, or a shape expression, and returns its canonical
form. This algorithm is recursive over the syntax of shapes and shape expressions. It simplifies
shapes, by computing operators of any constants, ordering variables by name inside arithmetic
operators, removing empty shapes in shape appends, and computing shape appends where possible.


\subsection*{Arithmetic Simplification}
These rules simplify arithmetic expressions.
\begin{align*}
  C \cup \{ \term{(*} \enspace d \enspace c_1 \term{)} = c_2 \}
  & \Rightarrow C \cup \{ s = \frac{c_2}{c_1} \} \enspace \text{if} \enspace c_2 \bmod c_1 = 0 \\
  C \cup \{ \term{(+} \enspace d \enspace c_1 \term{)} = c_2 \}
  & \Rightarrow C \cup \{ s = c_2 - c_1 \} \enspace \text{if} \enspace c_2 - c_1 > 0 \\
  C \cup \{ \term{(-} \enspace d \enspace c_1 \term{)} = c_2 \}
  & \Rightarrow C \cup \{ s = c_1 + c_2 \}
\end{align*}

%
%
%

\subsection*{Partial Expression Simplification}
This rule allows unsolvable expressions that equate to a constant to be replaced within other
constraints.
\begin{align*}
  & C + \{ (\term{op} \enspace s_1 \enspace s_2) = c \} \\
  & \Rightarrow C[c/(\term{op} \enspace s_1 \enspace s_2)] +
  \{ (\term{op} \enspace s_1 \enspace s_2) = c \} \\
  & \qquad \text{if} \enspace (\term{op} \enspace s1 \enspace s2) \in C
\end{align*}

\section{Inference Examples}
\label{sec:examples}

This section presents a set of example programs, and how shape inference can determine the shapes of
their inputs and outputs, based on both shapes provided by the programmer and the structure of the
program itself.

\subsection{Basic Symbolic Constraints}

The following Axon function performs matrix multiplication on a pair of square matrices of size
$n$. The inputs have shape $[n,n]$ and the return type is given name $r$. This example demonstrates
how the output shape is inferred from the input shape, and the structure of the program:
\begin{lstlisting}
fn (x : f32 [n,n], y : f32 [n,n]) : r {
  matmul(x, y)
}
\end{lstlisting}

The program uses the generalized matrix multiplication operator $\term{matmul}$, whose type
specification is as follows, where $\tau$ denotes the element type of the tensor:
\begin{equation*}
  (\tau \enspace s \term{@} \term{[} d_1 \term{,} d_2 \term{]},
  \enspace \tau \enspace s \term{@} \term{[} d_2 \term{,} d_3 \term{]}) \rightarrow \enspace
  \tau \enspace s \term{@} \term{[} d_1 \term{,} d_3 \term{]}
\end{equation*}

The set of shape constraints, generated by type checking this program, are:
\begin{align*}
  s \term{@} \term{[} d_1 \term{,} d_2 \term{]} & = \term{[} n \term{,} n \term{]} \\
  s \term{@} \term{[} d_2 \term{,} d_3 \term{]} & = \term{[} n \term{,} n \term{]} \\
  s \term{@} \term{[} d_1 \term{,} d_3 \term{]} & = r
\end{align*}

The first two constraints match the empty shapes rule, which assigns $s$ to be the empty shape. The
constraints are now:
\begin{equation*}
\begin{aligned}[c]
  \term{[} d_1 \term{,} d_2 \term{]} & = \term{[} n \term{,} n \term{]} \\
  \term{[} d_2 \term{,} d_3 \term{]} & = \term{[} n \term{,} n \term{]}
\end{aligned}
\quad
\begin{aligned}[c]
  s \term{@} \term{[} d_1 \term{,} d_3 \term{]} & = r \\
  s & = \term{[} \term{]}
\end{aligned}
\end{equation*}

The unpacking rule for shapes of known rank now applies to the first two constraints, giving:
\begin{equation*}
\begin{aligned}[c]
  d_1 & = n \\
  d_2 & = n \\
  d_3 & = n
\end{aligned}
\quad
\begin{aligned}[c]
  s \term{@} \term{[} d_1 \term{,} d_3 \term{]} & = r \\
  s & = \term{[} \term{]}
\end{aligned}
\end{equation*}

The reordering rule matches the fourth constraint, giving:
\begin{equation*}
\begin{aligned}[c]
  d_1 & = n \\
  d_2 & = n \\
  d_3 & = n
\end{aligned}
\quad
\begin{aligned}[c]
  r & = s \term{@} \term{[} d_1 \term{,} d_3 \term{]} \\
  s & = \term{[}\term{]}
\end{aligned}
\end{equation*}

The basic rule now matches constraint five, replacing $s$ with the empty shape in constraint four:
\begin{equation*}
\begin{aligned}[c]
  d_1 & = n \\
  d_2 & = n \\
  d_3 & = n
\end{aligned}
\quad
\begin{aligned}[c]
  r & = \term{[} \term{]} \term{@} \term{[} d_1 \term{,} d_3 \term{]} \\
  s & = \term{[} \term{]}
\end{aligned}
\end{equation*}

The basic rule also matches constraints one and three, replacing $d_1$ and $d_3$ in constraint four:
\begin{equation*}
\begin{aligned}[c]
  d_1 & = n \\
  d_2 & = n \\
  d_3 & = n
\end{aligned}
\quad
\begin{aligned}[c]
  r & = \term{[} \term{]} \term{@} \term{[} n \term{,} n \term{]} \\
  s & = \term{[} \term{]}
\end{aligned}
\end{equation*}

Shape simplification then removes the empty shape append in constraint four:
\begin{equation*}
\begin{aligned}[c]
  d_1 & = n \\
  d_2 & = n \\
  d_3 & = n
\end{aligned}
\quad
\begin{aligned}[c]
  r & = \term{[} n\term{,} n \term{]} \\
  s & = \term{[} \term{]}
\end{aligned}
\end{equation*}

Shape inference is now complete, as no rules match any of the constraints. The return shape has been
correctly inferred as $\term{[}n\term{,}n\term{]}$. Furthermore, the remaining constraints form a
shape substitution, mapping variables to shapes. The shapes for the arguments and return from the
$\term{matmul}$ call can then be found by applying this substitution to the type of
$\term{matmul}$):
\begin{equation*}
  (\tau \term{[} n \term{,} n \term{]}, \tau \term{[} n \term{,} n \term{]}) \rightarrow \tau \term{[} n \term{,} n \term{]}
\end{equation*}

\subsection{Arithmetic Expressions for Shapes}

The following program performs a 2d convolution of an 8x8 filter over an input of unknown shape
$i$. The result shape is known. This example demonstrates that shape inference can infer the shape
of the input from the shape of the output.
\begin{lstlisting}
fn (x : t i,
    f : t [4,8,8,8]
): t [4,8,1024,256] {
  conv(x, f)
}
\end{lstlisting}

The set of type constraints generated by type checking is:
\begin{equation*}
\begin{aligned}[c]
  \term{[} n \term{,} c \term{,} h \term{,} w \term{]} & = i \\
  \term{[} k \term{,} c \term{,} r \term{,} s \term{]} & = \term{[}4\term{,}8\term{,}8\term{,}8\term{]} \\
  \term{[} n \term{,} c \term{,} (\term{+} \enspace 1 \enspace (\term{-} \enspace h \enspace r))\term{,}
            (\term{+} \enspace 1 \enspace (\term{-} \enspace w \enspace s))\term{]}
  & = \term{[}4\term{,}8\term{,}1024\term{,}256\term{]}
\end{aligned}
\end{equation*}

The second and third constraints matches the unpacking rule for shapes of known rank, which gives
the following set of constraints:
\begin{equation*}
\begin{aligned}[c]
  \term{[} n \term{,} c \term{,} h \term{,}  w \term{]} & = i \\
  n & = 4 \\
  k & = 4 \\
  c & = 8
\end{aligned}
\quad
\begin{aligned}[c]
  r & = 8 \\
  s & = 8 \\
  (\term{+} \enspace 1 \enspace (\term{-} \enspace h \enspace r)) & = 1024 \\
  (\term{+} \enspace 1 \enspace (\term{-} \enspace w \enspace s)) & = 256
\end{aligned}
\end{equation*}

The basic rule for variable replacement then produces:
\begin{equation*}
\begin{aligned}[c]
  \term{[} 4 \term{,} 8 \term{,} h \term{,}  w \term{]} & = i \\
  n & = 4 \\
  k & = 4 \\
  c & = 8
\end{aligned}
\quad
\begin{aligned}[c]
  r & = 8 \\
  s & = 8 \\
  (\term{+} \enspace 1 \enspace (\term{-} \enspace h \enspace 8)) & = 1024 \\
  (\term{+} \enspace 1 \enspace (\term{-} \enspace w \enspace 8)) & = 256
\end{aligned}
\end{equation*}

Reordering reorders the first constraint:
\begin{equation*}
\begin{aligned}[c]
  i & = \term{[} 4 \term{,} 8 \term{,} h \term{,} w \term{]} \\
  n & = 4 \\
  k & = 4 \\
  c & = 8
\end{aligned}
\quad
\begin{aligned}[c]
  r & = 8 \\
  s & = 8 \\
  (\term{+} \enspace 1 \enspace (\term{-} \enspace h \enspace 8)) & = 1024 \\
  (\term{+} \enspace 1 \enspace (\term{-} \enspace w \enspace 8)) & = 256
\end{aligned}
\end{equation*}

The simplification algorithm applied to the arithmetic expressions reorders them into canonical
form:
\begin{equation*}
\begin{aligned}[c]
  i & = \term{[} 4 \term{,} 8 \term{,} h \term{,} w \term{]} \\
  n & = 4 \\
  k & = 4 \\
  c & = 8
\end{aligned}
\quad
\begin{aligned}[c]
  r & = 8 \\
  s & = 8 \\
  (\term{+} \enspace (\term{-} \enspace h \enspace 8) \enspace 1) & = 1024 \\
  (\term{+} \enspace (\term{-} \enspace w \enspace 8) \enspace 1) & = 256
\end{aligned}
\end{equation*}

The arithmetic simplification rule then produces:
\begin{equation*}
\begin{aligned}[c]
  i & = \term{[} 4 \term{,} 8 \term{,} h \term{,} w \term{]} \\
  n & = 4 \\
  k & = 4 \\
  c & = 8
\end{aligned}
\quad
\begin{aligned}[c]
  r & = 8 \\
  s & = 8 \\
  (\term{-} \enspace h \enspace 8) & = 1023 \\
  (\term{-} \enspace w \enspace 8) & = 255
\end{aligned}
\end{equation*}

And again:
\begin{equation*}
\begin{aligned}[c]
  i & = \term{[} 4 \term{,} 8 \term{,} h \term{,} w \term{]} \\
  n & = 4 \\
  k & = 4 \\
  c & = 8
\end{aligned}
\quad
\begin{aligned}[c]
  r & = 8 \\
  s & = 8 \\
  h & = 1031 \\
  w & = 263
\end{aligned}
\end{equation*}

The basic rule for variable replacement then produces:
\begin{equation*}
\begin{aligned}[c]
  i & = \term{[} 4 \term{,} 8 \term{,} 1031 \term{,} 263 \term{]} \\
  n & = 4 \\
  k & = 4 \\
  c & = 8
\end{aligned}
\quad
\begin{aligned}[c]
  r & = 8 \\
  s & = 8 \\
  h & = 1031 \\
  w & = 263
\end{aligned}
\end{equation*}

And shape inference is complete. The shape of the input has been inferred as $\term{[}4\term{,}8\term{,}1031\term{,}263\term{]}$.

\subsection{Discovering Shapes of Inputs}

The shape inference algorithm allows the shapes of unknown inputs to be inferred from other
information in the program. For example, consider the following example. The type signature of
\lstinline|max| determines the shape of the second input, given the shape of the first input. This
demonstrates that shape information can be inferred not only ``forward'' from input shapes to output
shapes, but also ``in reverse'' back to the inputs.
\begin{lstlisting}
fn(a : f32, b) {
  max(a, b)
}
\end{lstlisting}

Here the shape of \lstinline|b| is inferred to also be a scalar \term{f32}, due to the other input
being a scalar \term{f32} and the type signature for \lstinline{max}. The return type of the
function is also inferred to be \term{f32}.

\subsection{Partial Shape Information}

Sometimes there is not enough information to determine concrete shapes for all parts of a
program. Our shape inference algorithms allows partial shape information to be discovered, for
example consider the following \lstinline|softmax| program:
\begin{lstlisting}
softmax = fn (x : f32 s) {
  y = reduce(max, -inf, x);
  z = exp(x - y);
  z / reduce(+, 0f, z)
}
\end{lstlisting}

No shape information is given in the type of the input (it is just denoted \lstinline|s|), however
type information can be determined from the operations contained within the program. Running shape
inference on this program yields the following type for the \lstinline|softmax| function:
\begin{equation*}
\term{f32} \enspace \term{[}a\term{]} \term{@} b  \enspace \rightarrow \enspace \term{f32} \enspace \term{[}a\term{]} \term{@} b
\end{equation*}

Shape inference has discovered, based on the type signatures of the operations within the program,
that it can accept any input with a rank of at least one (the first dimension is denoted $a$) and
arbitrary outer dimensions (denoted $b$). Furthermore, the output of the function is also inferred
to have the same shape as the input.


\section{Example Graphs}
\label{sec:example-graphs}

This section presents example graphs, and how shape inference can determine the shapes of
their inputs and outputs.

\subsection{Attention}

A commonly used computation in machine learning models is attention \cite{Vaswani2017Attention}.
This can be written as the following mathematical function:

\begin{equation*}
\text{attention}(Q,K,V) = \text{softmax}(\frac{QK^T}{\sqrt{d_k}}V)
\end{equation*}

This can then be implemented in Axon as follows:

\begin{lstlisting}
softmax = fn (e : t [m,n]) {
  z = exp(e);
  s = map(
    fn (a) {
      reduce (+, 0f, a)
    },
    transpose(e)
  );
  map (fn(a) { a / s }, z)
};

attention = fn(Q, K, V) {
  matmul(softmax(matmul(Q,transpose(K))), V)
};
\end{lstlisting}

In this program, all of the types and shapes are implicit and unconstrained, except for the input to
the softmax function, which has type $t \term{[} m \term{,} n \term{]}$. This type annotation
constrains the input to be of any element types, with a shape which is two dimensional, with
dimensions $m$ and $n$ respectively.

Axon includes rules for generating shape constraints for all of the built-in functions used in this
program, such as matmul, transpose and map as described in
Section~\ref{sec:example-shape-constraints}. These rules generate a set of constraints, which when
solved give the following type and shape signature for the attention function:
\begin{equation*}
  t \term{[} a \term{,} b \term{]} \term{,} \enspace t \term{[} c \term{,} b \term{]}
  \term{,} \enspace t \term{[} c \term{,} d \term{]} \enspace \rightarrow \enspace t \term{[} a
    \term{,} d \term{]}
\end{equation*}

\subsection{Loops and Scans}

Loops and scans, as used in LSTM models \cite{LSTM}, can be written in Axon as follows:

\begin{lstlisting}
h_out, y = loop(h_in, cell, x)
\end{lstlisting}

where \lstinline|cell| is a function that takes the current state and outputs the next state,
\lstinline|x| is the input tensor, \lstinline|h_in| is the initial hidden state, \lstinline|h_out|
is the final hidden state and \lstinline|y| is the output tensor.

This loop construct can be used to write an LSTM as follows, by simply specifying the appropriate
cell function for the LSTM:

\begin{lstlisting}
cell = fn (state : (t s1, t s2), y : t s3) {
  h1, c1 = state;
  hnext, cnext = lstmStep(
    y, h1, c1,
    Wi, Wf, Wo, Wc,
    Ri, Rf, Ro, Rc,
    Bi, Bf, Bo, Bc);
  (hnext, cnext), cnext
};
loop((h, c), cell, x)
\end{lstlisting}

As with the attention example, most of this program is left with unconstrained types and shapes. The
only requirement is that the inputs to the cell function are a tuple of two tensors, and a
tensor. In this example the shape solver will determine that the output of the loop has the same
shape as the initial input tensor x, and given the shape of x, constrains the cell function to have
the correct shapes. Namely, that $s1$, $s2$ and $s3$ should be equal to the shape of x.

\subsection{Recurrent Neural Networks}

Bidirectional recurrent neural networks can be written in Axon as follows:

\begin{lstlisting}
bidir = fn (a, b, c, d, s) {
    _, x = loop(a, b, s);
    _, y = loop(c, d, reverse(s));
    concat(x, reverse(y))
};
Y = bidir(s0, A, s0', A', X);
\end{lstlisting}

This uses the same loop construct as the LSTM example previously.

Similarly, residual recurrent neural networks can be written in Axon as follows:

\begin{lstlisting}
residual = fn (a, b, s) {
  _, r = loop(a, b, s);
  r + s
};
X1 = residual(..., LSTM1, X0);
X2 = residual(..., LSTM2, X1);
\end{lstlisting}

\pagebreak

\section{Related Work}
\label{sec:related-work}

Roesch~et~al.~\cite{Roesch2018RelayAN} describe RelayIR -- an intermediate representation for deep
learning. Similarly to Axon, RelayIR is a functional statically typed language aimed at deep
learning problems. In contrast to our work, it uses a dependent type system to allow shapes to be
represented by expressions. Our approach instead extends the language of shapes to include
arithmetic expressions, shape appending and other constructs in order to allow complex shapes to be
represented.

As far as we are aware there are no other related works for deep learning in the literature, that
use a constraint based methodology to infer shape information in graph programs.

Frameworks, such as TensorFlow~\cite{Tensorflow2025Whitepaper} and Pytorch~\cite{PytorchNIPS2019}
can propagate shape information at graph construction time but do not use an approach that allows
shape information to be inferred from outputs to inputs.

The MAGICA project~\cite{Joisha2006MATLAB} explores shape inference for MATLAB programs, using a symbolic
evaluation approach, based on the algebraic properties of MATLAB. They use it to identify redundant
shape array checks in MATLAB programs in order to reduce execution overheads at runtime.

Garg~et~al.~\cite{GargARRAY19} describe an approach that uses runtime information to perform shape
inference on APL-like languages for just-in-time compilation. In contrast, our inference algorithm
can be applied both statically at compile time, to infer potentially partial shape information, and
again at runtime (if required) once the shapes of graph inputs are known.

\section{Conclusions}
\label{sec:conclusions}

This paper presents an algorithm for generating and solving shape constraints for Deep Learning
graphs with dynamic shapes. This algorithm allows either fully static or partially static shapes to
be discovered within a program, based on shape information provided by the programmer, and the
structure of the graph. These shapes can include arithmetic expressions for their individual
dimensions, and can also vary in rank. This is done using a fully automated shape constraint solver.

\pagebreak

%
%
%

\bibliography{paper}

%

\end{document}